\documentstyle [12pt]{article}
\begin{document}

\underline{Are Neutrino Oscillations Compatible with the Decaying
Neutrino}
 
\underline{Theory for the Ionisation of Hydrogen in the Interstellar
Medium?}

\bigskip

\centerline{D.W. Sciama{*}}

\bigskip

SISSA and ICTP, Strada Costiera 11, 34014 Trieste, Italy

Department of Physics, Oxford University\\

Recent Super-Kamiokande data have greatly strengthened the hypothesis
that the atmospheric neutrino anomaly is mainly due to nearly maximal
oscillations between $\nu_{\mu}$ and either $\nu_{\tau}$ or a sterile 
neutrino $\nu^s_{\mu}$.In this letter we point out that the decaying
neutrino 
theory for the ionisation of hydrogen in the interstellar medium, in
conjunction with the assumption that the cosmological constant $\lambda$
is zero, requires these oscillations to be between $\nu_{\mu}$ and
$\nu^s_{\mu}$. This requirement can be tested by forthcoming
Super-Kamiokande
measurements.

A further requirement can be derived if the solar neutrino deficit is
energy 
dependent and if both it and
the LSND data are also due to neutrino oscillations.  In that case
the decaying neutrino theory alone (without any assumption about the
value of $\lambda$) requires both that the atmospheric anomaly is due to
$\nu_{\mu}-\nu^s_{\mu}$ oscillations and that the solar neutrino
oscillations
are between $\nu_e$ and a second sterile neutrino $\nu^s_e$.  This
latter requirement can be tested by forthcoming SNO and BOREXINO
measurements.

If these are also $\nu_{\tau}-\nu^s_{\tau}$ oscillations the decay line
would be split into two lines of nearly equal intensity and spacing less
than  $\sim$ 0.2 eV.

\newpage

Recent Super-Kamiokande data [1] have greatly strengthened the
hypothesis that the observed deficit in the flux of atmospheric
$\nu_{\mu}$'s is mainly due to their nearly maximal oscillation into
another
neutrino flavor, which a priori could be $\nu_e, \nu_{\tau}$ or
$\nu^s_{\mu}$
(a hypothetical sterile neutrino which lacks the standard electroweak
interactions).  These and other relevant neutrino data have been
analysed in detail [2] with the following results:
\medskip

\hangindent 1.5 truecm
(i) $\nu_{\mu}-\nu_e$ oscillations can be ruled out
as the main origin of the
deficit
\medskip

\hangindent 1.5 truecm
(ii) $\nu_{\mu}-\nu_{\tau}$ and $\nu_{\mu}-\nu^s_{\mu}$
oscillations fit the
data nearly equally well, with $\delta$m$^2 \sim 10^{-3} - 10^{-2} eV^2$.
\medskip

\hangindent 1.5 truecm
(iii) forthcoming data should be able to resolve
this ambiguity
relatively soon.  This resolution is possible because $\nu_{\tau}$ has
neutral current interactions with matter in the detector and in the
Earth, while $\nu^s_{\mu}$ does not.  This difference affects both the
production rate of neutral pions in the detector, and the energy and
zenith angle dependence of the $\nu_{\mu}$ flux.
\medskip

The $\nu_{\mu}-\nu^s_{\mu}$ option has also to pass the test of not
violating
constraints from big bang nucleosynthesis (BBNS).  It used to be argued
[3] that maximal mixing between $\nu_{\mu}$ and $\nu^s_{\mu}$ would give
rise to
a
thermal distribution of $\nu^s_{\mu}$ in the early universe, leading to an
effective number N$_\nu$ of neutrino flavors equal to 4.  This would
contradict the implications of the observed abundances of the light
elements, especially D and He$^4$.  However, it has recently been shown
[4] that $\nu_{\mu}-\nu^s_{\mu}$ oscillations in the early universe would
induce a lepton asymmetry which in turn could lead to a significant
reduction in N$_\nu$ from the expected value of 4.  Accordingly the
$\nu{_\mu}-\nu^s_{\mu}$ option remains a viable one for resolving the
atmospheric neutrino anomaly.

In this letter we point out that the decaying neutrino theory [5] for the
origin of the ionisation of hydrogen in the interstellar medium of our
Galaxy could resolve the ambiguity between $\nu_{\tau}$ and $\nu^s_{\mu}$,
in
favor of $\nu^s_{\mu}$, if the cosmological constant $\lambda$ is zero.
According to this theory dark matter neutrinos decay into hydrogen -
ionising photons and less massive neutrinos.  In order for these decay
photons to be the main source of the free electrons in the interstellar
medium the decay lifetime needs to be $\sim 2 \times 10^{23}$ sec.  The
mass m$_1$ of the decaying neutrino is constrained [6] to be 27.4 $\pm$
0.2 eV, if the mass of the secondary neutrino in the decay $\ll$ m$_1$.
It would then follow [6] from this that $\Omega \rm h^2 \sim 0.3$, where
$\Omega$ is the density of the universe in units of the critical density
and h is the Hubble constant in units of 100 km.sec$^{-1}$ Mpc$^{-1}$.

Some of the recent observational estimates of h [7] and of the age t$_0$
of the universe [8], while still uncertain, would be compatible with our
constrained value of $\Omega$ h$^2$ if $\Omega$ = 1 (so that h = 0.55
and t$_0$ = 12 Gyr).  This compatibility is achieved for $\lambda$ = 0,
which we regard as an attractive feature of the decaying neutrino theory
that we would wish to maintain for the time being, despite the
indications from recent observations of supernovae at large red shifts
[9] that $\Omega$ $<$ 1 and $\lambda$ $>$ 0.  This evidence is
preliminary and needs to be confirmed.

On the other hand there does exist some evidence [10] in favor of the
decaying neutrino theory, based on Hubble Space Telescope observations
[11] of the spectrum of the halo star HD93521.  This suggests that it
would be worthwhile to explore the implications of this theory for
phenomena involving neutrino oscillations.  In particular we shall now
show that it would resolve the ambiguity between $\nu_{\tau}$ and
$\nu^s_{\mu}$ 
in the atmospheric neutrino anomaly if $\lambda$ = 0.  To see this we
note that, since m$_{\nu_{e}} <$ 15 ev [12], the decaying neutrino must
be either $\nu_{\mu}$ or $\nu_{\tau}$.  Since $\delta m^2$ is so small,
if $\nu_{\mu}$ oscillations into $\nu_{\tau}$ were the origin of the
atmospheric
anomaly, it would follow that m$_{\nu_{\mu}} \sim m_{\nu_{\tau}}$.
Accordingly the density of the universe would have to be nearly doubled
from our previous value, so that $\Omega$ h$^2$ $\sim$ 0.6.  If one
assumes that $\lambda$ = 0 and t$_o$ $\geq$ 12 Gyr, one would require
h $\leq$ 0.4 and $\Omega$ $\geq$ 3.7, values which are strongly
disfavored.  Accordingly one concludes that the atmospheric neutrino
anomaly must be due to nearly maximal $\nu_\mu - \nu^s_{\mu}$ 
oscillations.

This prediction of the decaying neutrino theory should soon be tested by
the forthcoming experiments mentioned earlier.  It also may have
astrophysical implications since oscillations of $\nu_{\mu}$
into $\nu^s_{\mu}$ may play a crucial role in r-process nucleosynthesis
and shock revival in supernova envelopes [13], in the production of
highly relativistic jets in $\gamma$ ray burst events [14], and in 
baryosynthesis [15].

We can make a further deduction from the decaying neutrino theory
(without any assumption about the value of $\lambda$) if the solar
neutrino [16] and Los Alamos
LSND [17] anomalies are also due to neutrino oscillations.  The basic
results of the latter experiment are still subject to considerable
uncertainty, which should, however, be resolved by forthcoming
experiments.  For the present discussion, and following many authors [2],
we shall provisionally assume that the LSND data are correct. If it turns
out that they are not, we can still use the assumption that solar neutrino
oscillations are occurring to deduce that the atmospheric neutrino
oscillations involve $\nu_{\mu}$ and $\nu^s_{\mu}$ even if $\lambda \ne
0$.

It has been widely pointed out [2] that, if the solar neutrino deficit 
is energy dependent, the $\delta$ m$^2$'s implied by
the three oscillation experiments are so disparate that it is necessary
to invoke at least 4 neutrino flavors in order to accommodate the 3
different $\delta$ m$^2$'s.  In the literature these 4 flavors are
usually assumed to be $\nu_e, \nu_{\mu}, \nu_{\tau}$ and $\nu^s$ (since a
fourth active neutrino is forbidden by the LEP data [18] on decaying Z
particles).  It has been argued [19] that the totality of neutrino
oscillation data are best fit by a neutrino mass spectrum which consists
of two nearly degenerate pairs of mass states, one pair being
responsible for the solar neutrino anomaly and the other for the
atmospheric anomaly.  In this scheme, either the solar anomaly would be
due to $\nu_e-\nu^s_e$ oscillations and the atmospheric anomaly to
$\nu_{\mu}- \nu_{\tau}$ oscillations, or the solar anomaly to $\nu_e-
\nu_{\tau}$ and the atmospheric one to $\nu_{\mu}-\nu^s_{\mu}$.

In addition the two nearly degenerate pairs of mass states would be
separated by $\sim$ 1 ev to accommodate the LSND data.  This mass spectrum
has also been used to provide for the cosmological hot dark matter in the
mixed hot and cold dark matter scenario, where $\Sigma m_{\nu} \sim$ 2 ev
[20].  By contrast, in the decaying neutrino theory m$_{\nu_{\tau}}$ would
be much too large for $\nu_{\tau}$ to belong to the 4-neutrino spectrum,
and $\nu_{\mu}-\nu_{\tau}$ oscillations could be ruled out for the
atmospheric anomaly even if $\lambda \neq 0$. However, this spectrum, and
its compatibility with the data, can be retained in our theory if a second
sterile neutrino $\nu^s_e$ is introduced to act as a partner to
$\nu_e$.  Then the solar neutrino anomaly would be mainly due to
$\nu_e-\nu^s_e$ oscillations rather than $\nu_e-\nu_{\mu}$ or
$\nu_e-\nu_{\tau}$. Models with two and with three sterile neutrinos have,
in fact, already been discussed in the literature [21]. According to the
most recent analysis [16] of the solar neutrino data, the
$\nu_e-\nu^s_e$ solution is a viable one.
Our prediction that the solar neutrino anomaly is mainly due to
$\nu_e-\nu^s_e$ oscillations can be tested [16] in forthcoming
measurements by SNO and BOREXINO, for example by measuring the $\nu-e$
scattering rate for the 0.86 MeV Be$^7$ line. 

Finally we consider the possibility that there is a third sterile neutrino
$\nu^s_{\tau}$ which is associated with $\nu_{\tau}$. A particularly
interesting situation would arise if $\nu_{\tau}$ and $\nu^s_{\tau}$ were
maximally mixed and if the mass eigenstates were split by an amount small
compared to each eigenmass. The $\nu_{\tau}-\nu^s_{\tau}$ system would
then
constitute a pseudo-Dirac neutrino [23]. To avoid a resultant doubling of
the density of the universe we appeal to the Foot-Volkas mechanism [4] to
suppress the sterile neutrinos which would otherwise be produced during
the era in the early universe in which $\nu_{\tau}$ is coupled to the heat
bath. To confirm this possibility this mechanism would need to be extended
to the case where all three active neutrinos are oscillating into their
sterile partners in the early universe.

After $\nu_{\tau}$ has decoupled its abundance would be depleted by its
oscillations into $\nu^s_{\tau}$. In the present universe we would expect
to find equal number densities ($\sim$ 55 cm$^{-3}$) of
$\nu_{\tau}$ and $\nu^s_{\tau}$. We now consider the effect of this
situation on the decaying neutrino theory. Each mass eigenstate consists
of a superposition of active and sterile components with equal amplitudes
(because the mixing has been assumed to be maximal). Then each mass
eigenstate would radiatively decay via its active component with nearly
the same lifetime (since the mass splitting has been assumed to be
small). The decay line would thus be split into two narrowly separated
lines with nearly the same intensity. (There would also be a further
small
splitting due to the two mass eigenstates of the secondary neutrino in the
decay).

We can place constraints on the extent of this line splitting from
astronomical observations. The harder line cannot be too hard or the
observed upper limit on the H$\alpha$ flux from intergalactic neutral
hydrogen clouds would be exceeded [5]. The softer line cannot have an
energy less than 13.6 eV because if it did, the dark matter neutrinos in
the Galaxy would produce an unabsorbed line of enormous intensity which
would have already been easily observed [24].
These considerations limit each of the lines to lying above the Lyman
limit with the sum of their excesses being bounded by about 0.2 eV.
 
If the line splitting could be resolved by a suitable detector each line
would have half the intensity of the previously derived line, since the
decay
lifetime would be nearly the same for each mass eigenstate as the mass
splitting is small.

This analysis suggests that all three active neutrinos might be of the
pseudo-Dirac type, as has already been proposed by Geiser, by Bowes and
Volkas and by Koide and Fusaoka (see (21)). The implied maximal mixing of
$\nu_e$ and $\nu^s_e$  could be compatible with the MSW explanation
[22] of the solar neutrino data if the published error bars on these data
were increased by a factor 1.7 for all four experiments (Homestake,
GALLEX, SAGE and Superkamiokande) or by less than 2.5 for the Homestake
experiment alone [25]. 

We conclude that if the decaying neutrino theory for the ionisation of
hydrogen in the interstellar medium is correct, and if the cosmological
constant $\lambda$ is zero, then the atmospheric neutrino data must
involve $\nu_{\mu}-\nu^s_{\mu}$ oscillations.  If, in addition, the solar
neutrino deficit is energy dependent, and both it and the LSND data are
due to neutrino oscillations (but $\lambda$ is not necessarily zero), then
the solar neutrino oscillations must involve $\nu_e$ and $\nu^s_e$. 
Both these conclusions can be tested by forthcoming experiments. 

If, in addition, there are $\nu_{\tau}-\nu^s_{\tau}$ oscillations, then
the
decay line would be split into two lines of equal intensity and spacing
less than 0.2 eV. This splitting might be observable.

These possibilies are compatible with all three active neutrinos being of
the pseudo-Dirac type if the published error bars of the solar neutrino
experiments are somewhat increased.
 
I am grateful to E.K. Akhmedov, B.C. Allanach, R. Foot, Q.Y. Liu, D.H.
Perkins, A.Masiero, S.T. Petcov, G.G. Ross, S. Sarkar, A.Y. Smirnov and
R.R.Volkas
for helpful discussions and the Italian MURST for their financial
support of this work.

\pagebreak

* Email address: sciama@sissa.it\\

\hangindent 1.5 truecm
[1] Y. Fukuda et al, Phys. Lett.B {\bf{433}}, 9 (1998); Phys. Rev. Lett. 
{\bf{81}}, 1562 (1998).

\medskip
\hangindent 1.5 truecm
[2] M.C. Gonzalez-Garcia et al, Phys. Rev. D {\bf{58}}, 033004 (1998);
hep-ph/9807305 S.C. Gibbons et al, Phys. Lett. B {\bf{430}}, 296 (1998) 7.
F.
Vissani and A.Y. Smirnov, Phys. Lett. B {\bf{432}}, 376 (1998); M.
Narayan,
G. Rajasekharan and S.V. Sanker, Phys. Rev. D. {\bf{58}}, 031301 (1998);
Q.Y. Liu and A.Y. Smirnov, Nucl. Phys. B {\bf{524}}, 505 (1998); R. Foot,
R.R. Volkas and O. Yasuda, Phys. Rev. D.{\bf{58}}, 013006 (1998); Phys.
Lett. B, {\bf{433}} 82
(1998); P. Lipari
and M. Lusignoli, hep-ph/9803440.; R.P. Thun and S.McKee, hep-ph/9806534.

\medskip
\hangindent 1.5 truecm
[3] R. Barbieri and A. Dolgov, Phys. Lett. B.
{\bf{237}}, 440 (1990); K.
Enquist, K. Kainulainen and J. Maalampi, Nucl. Phys. B. {\bf{349}}, 754
(1991); K. Enquist, K. Kainulainen and M. Thomson, Nucl. Phys. B.
{\bf{373}}, 498 (1992); J.M. Cline, Phys. Rev. Lett. {\bf{68}}, 3137
(1992);
X. Shi, D.N. Schramm and B.D. Fields, Phys. Rev. D. {\bf{48}}, 2563
(1993).

\medskip
\hangindent 1.5 truecm
[4] R. Foot and R.R. Volkas, Phys. Rev D {\bf{55}},
5147 (1997),
{\bf{56}}, 6653 (1997); astro-ph/9811067; S.M. Bilenky et al,
hep-ph/9804421.

\medskip
\hangindent 1.5 truecm
[5] D.W. Sciama, Ap.J. {\bf{364}}, 549 (1990), Modem
Cosmology and the
Dark Matter Problem (Cambridge; Cambridge University Press 1993).

\medskip
\hangindent 1.5 truecm
[6] D.W. Sciama, MNRAS {\bf{289}}, 945 (1997).

\medskip
\hangindent 1.5 truecm
[7] D. Branch, astro-ph/9801065; G.A. Tammann,
astro-ph/9805013. 
B.E. Schaefer, astro-ph/9808157.

\medskip
\hangindent 1.5 truecm
[8] R.G. Gratton et al, Ap.J. {\bf{491}}, 749
(1997); B.Chaboyer, astro-ph/9808200.

\medskip
\hangindent 1.5 truecm
[9] S. Perlmutter et al, Nature.{\bf{391}}, 51
(1998); 
A.V. Filippenko and A.G. Riess, astro-ph/9807008.

\medskip
\hangindent 1.5 truecm
[10] D.W. Sciama, Ap.J. {\bf{488}}, 234 (1997).

\medskip
\hangindent 1.5 truecm
[11] L. Spitzer and E.L. Fitzpatrick, Ap.J.
{\bf{409}}, 299 (1993).

\medskip
\hangindent 1.5 truecm
[12] Particle Physics Summary, Rev. Mod. Phys.
{\bf{68}}, 611 (1996).

\medskip
\hangindent 1.5 truecm
[13] J.T. Peltoniemi, hep-ph/9511323. Astron. and
Astrophys.{\bf{254}}, 
121 (1992) H.Nunokawa et al, Phys. Rev. D {\bf{56}}, 
1704 (1997); D.O. Caldwell, G.M. Fuller and Y.Z. Qian, in preparation.

\medskip
\hangindent 1.5 truecm
[14] W. Kluzniak, astro-ph/9807224.

\medskip
\hangindent 1.5 truecm
[15] E.K. Akhmedov, V.A. Rubakov and A.Y. Smirnov,
Phys. Rev. Lett. {\bf{81}}, 
1359 (1998).

\medskip
\hangindent 1.5 truecm
[16] J.N. Bahcall, P.I. Krastev and A.Y. Smirnov,
hep-ph/9807216.

\medskip
\hangindent 1.5 truecm
[17] C. Athanassopoulos et al, Phys. Rev. Lett.
{\bf{81}}, 1774 (1998).

\medskip
\hangindent 1.5 truecm
[18] P.B. Renton, Int. J. Mod. Phys. A. {\bf{12}},
4109 (1997).

\medskip
\hangindent 1.5 truecm
[19] N. Okuda and O. Yasuda, Int. J. Mod. Phys. A
{\bf{12}}, 
3669 (1997); S.M. Bilenky, C. Giunti and W. Grimus, Eur. Phys. J. C.
{\bf{1}},
247 (1998); V. Barger et al, hep-ph/9806328.

\medskip
\hangindent 1.5 truecm
[20] J.R. Primack et al, Phys. Rev. Lett. {\bf{74}},
2160 (1995).

\medskip
\hangindent 1.5 truecm
[21] R. Foot, Mod. Phys. Lett. A9, {\bf{169}} (1994); R. Foot and R.R.
Volkas, Phys. Rev. D {\bf{52}}, 6595 (1995); A. Geiser, CERN-EP/98-56;
J.P. Bowes and R.R. Volkas, hep-ph/9804310; D. Suematsu, 
hep-ph/9805305; Y. Koide and H. Fusaoka, hep-ph/9806516; W. Krolikowski,
hep-ph/9808207.

\medskip
\hangindent 1.5 truecm
[22] S.P. Mikheyev and A.Y. Smirnov, Nuovo.Cim.C
{\bf{9}}, 
17 (1986); L. Wolfenstein, Phys. Rev. D {\bf{17}}, 2369 (1978), D
{\bf{20}}, 
2634 (1979).

\medskip
\hangindent 1.5 truecm
[23] L. Wolfenstein, Nucl. Phys. B {\bf{186}}, 147 (1981); S.M. Bilenky
and S. T. Petcov, Rev. Mod. Phys. {\bf{59}}, 671 (1987).

\medskip
\hangindent 1.5 truecm
[24] J.B. Holberg, Ap.J., {\bf{311}}, 969 (1986); E.J. Korpela, S.
Bowyer and J. Edelstein, Ap.J. {\bf{495}}, 317 (1998). 

\medskip
\hangindent 1.5 truecm
[25] Q.Y. Liu (private communication).

\end{document}